# Anomalous Hall effect mechanisms in quasi-2D van der Waals ferromagnet Fe$_{0.29}$TaS$_2$

Ranran Cai[1,2], Wenyu Xing[1,2], Huibin Zhou[1,2], Boning Li[1,2], Yangyang Chen[1,2], Yunyan Yao[1,2], Yang Ma[1,2], X. C. Xie[1,2,3,4], Shuang Jia[1,2,3,4], and Wei Han[1,2*]

[1] International Center for Quantum Materials, School of Physics, Peking University, Beijing 100871, P. R. China

[2] Collaborative Innovation Center of Quantum Matter, Beijing 100871, P. R. China

[3] CAS Center for Excellence in Topological Quantum Computation, University of Chinese Academy of Sciences, Beijing 100190, P. R. China

[4] Beijing Academy of Quantum Information Sciences, Beijing 100193, P. R. China

[*] Correspondence to: weihan@pku.edu.cn

**Abstract**

The recent emergence of two-dimensional (2D) van der Waals ferromagnets has provided a new platform for exploring magnetism in the flatland and for designing 2D ferromagnet-based spintronics devices. Despite intensive studies, the anomalous Hall effect (AHE) mechanisms in 2D van der Waals ferromagnets have not been investigated yet. In this paper, we report the AHE mechanisms in quasi-2D van der Waals ferromagnet Fe$_{0.29}$TaS$_2$ via systematically measuring Fe$_{0.29}$TaS$_2$ devices with thickness from 14 nm to bulk single crystal. The AHE mechanisms are investigated via the scaling relationship between the anomalous Hall and channel conductivities. As the Fe$_{0.29}$TaS$_2$ thickness decreases, the major AHE mechanism changes from extrinsic scattering to intrinsic contribution. The crossover of the AHE mechanisms is found to be highly associated with the channel conductivities as the $Fe_{0.29}TaS_2$ thickness varies.

## I. INTRODUCTION

The identification of two-dimensional (2D) van der Waals ferromagnets, including $Cr_2Ge_2Te_6$ and $CrI_3$, has provided a new platform for investigating fundamental low-dimensional magnetism and for potential spintronics devices using their heterostructures [1-9]. Interestingly, Ising-type ferromagnetism in bilayer $CrI_3$ gives rise to the recent observation of giant tunneling magnetoresistance and electrical field control of magnetism [10-14]. The heterostructures based on 2D ferromagnetic materials have also been studied, such as the interface magnetic proximity effect and magnetic tunneling resistances [15,16]. Long-distance magnon transport has been demonstrated in the 2D van der Waals antiferromagnetic insulator $MnPS_3$ thin flakes [17]. Despite these intensive studies, the anomalous Hall effect (AHE) mechanisms, among the most important physical properties of ferromagnetic materials [18], of the 2D van der Waals ferromagnets have not been reported yet.

In this paper, we report the investigation of the AHE mechanisms in quais-2D van der Waals ferromagnet $Fe_{0.29}TaS_2$ via the scaling relationship between the anomalous Hall and channel conductivities. It is found that the major AHE mechanism changes from extrinsic scattering to intrinsic contribution as the $Fe_{0.29}TaS_2$ thickness decreases from bulk single crystal to 14-nm flake. A strong correlation between the AHE mechanisms and the channel conductivities is identified, and the crossover of the AHE mechanisms is observed with a critical channel conductivity of $\approx 8 \times 10^3 \ \Omega^{-1} \ cm^{-1}$. Furthermore, a strong thickness dependence of the coercive magnetic field ($B_c$) is observed; $B_c$ decreases as the thickness decreases. These results could pave the way for exploring the fundamental magnetic properties of the 2D van der Waals ferromagnets and might be extremely useful for designing 2D materials-based spintronics devices, such as magnetic tunneling junctions and magnetic domain-wall racetrack memories [19,20].

## II. EXPERIMENTAL

Figure 1(a) shows the crystalline structure of the layered Fe-intercalated van der Waals ferromagnet $Fe_xTaS_2$, where the Fe atoms are located between the two $TaS_2$ layers. These layers are stacked together via van der Waals interaction with an interlayer distance of $\approx$ 6 Å. $Fe_xTaS_2$ single crystals are synthesized using the iodine vapor transport method, as described in previous reports [16,21-24]. The concentration of Fe ($x$) in synthesized $Fe_xTaS_2$ single crystals is determined to be 0.29 by energy-dispersive spectroscopy (Fig. S1 in Supplemental Material [25]). The crystalline property of synthesized $Fe_{0.29}TaS_2$ single crystals is characterized by high-resolution x-ray diffraction. As shown in Fig. 1(b), only sharp (002) and (004) $Fe_xTaS_2$ peaks are observed. $Fe_{0.29}TaS_2$ thin flakes are prepared on the $SiO_2$ (300 nm)/Si substrates using the well-established mechanical exfoliation method [26].

The $Fe_{0.29}TaS_2$ thin flakes are firstly identified using a Nikon high-resolution optical microscope, and their thicknesses are determined via atomic force microscopy. To perform the AHE measurements, standard Hall bar devices are fabricated on $Fe_{0.29}TaS_2$ thin flakes via electron-beam lithography, and the electrodes of ~80 nm Pt are deposited by rf sputtering in a magnetron sputtering system with a base pressure lower than $5 \times 10^{-5}$ torr. The stability of $Fe_{0.29}TaS_2$ thin flakes is characterized by Raman spectra. The almost identical Raman results measured on the pristine $Fe_{0.29}TaS_2$ flakes ($\approx$ 15 and 25 nm) and the devices (Fig. S2 in Supplemental Material [25]) indicate the negligible role of oxidation. This observation is consistent with a previous report that little oxidation effect is observed on $Fe_{0.25}TaS_2$ characterized by transmission electron microscopy [16]. Then the transport measurements are carried out in a physical property measurement system (Quantum Design).

**III. RESULTS**

Figures. 1(c-d) show the Hall resistances ($R_{xy}$) vs the perpendicular magnetic field ($B$) measured on a typical 14-nm $Fe_{0.29}TaS_2$ device at 10, 30, 50, 70, and 80 K, respectively. Fig. 1(c) inset shows the optical image of this device, where the thickness ($t$) of $Fe_{0.29}TaS_2$ flake is determined by the line profiles via atomic force microscopy (Fig. S3 in Ref. [25]). These square-like AHE curves indicate the strong out-of-plane magnetic anisotropy of the $Fe_{0.29}TaS_2$ thin flake, which is consistent with previous study reporting Ising-like spin states in bulk $Fe_xTaS_2$ single crystals [24]. At relative high magnetic fields, the negative sign of the slope for $R_{xy}$ vs $B$ indicates that the carriers are hole-type. The AHE curves loses the square-like shape at $T$ = 70 K and disappears at $T$ = 80 K (Fig. 1(d)), indicating that the Curie temperature is ≈ 80 K. The temperature-dependent anomalous Hall resistance ($R_{AHE}$) and channel resistance ($R_{xx}$) of this 14-nm $Fe_{0.29}TaS_2$ device are summarized in Figs. 1(e-f) and the detail analysis procedures are shown in Fig. S4 in Supplemental Material [25]. As the temperature increases, the enhancement of $R_{xx}$ is associated with stronger phonon and magnon scattering.

Interestingly, $B_c$ is only ≈ 0.04 T at $T$ = 10 K for this 14-nm $Fe_{0.29}TaS_2$ device (Fig. 2(a)), which is significantly small compared to $B_c$ of the bulk $Fe_xTaS_2$ [21-24]. To investigate the underlying mechanism for the variation of $B_c$, we perform the AHE measurements on the devices fabricated on $Fe_{0.29}TaS_2$ thin flakes with thickness from 14 nm to bulk single crystal. Fig. 2(b-c) show the AHE results at $T = 10$ K for devices made on 42-nm thin flake and bulk single crystal, respectively. An obvious trend is observed that $B_c$ increases dramatically as the $Fe_{0.29}TaS_2$ thickness increases, and it reaches $\approx 2.6$ T for bulk single crystal. Figures 2(d) shows the temperature dependence of $B_c$ for all the devices and Fig. 2(e) summarizes the thickness dependence of $B_c$ at $T = 10$ K. To our best understanding, this observation might be associated with thickness-dependent ferromagnetic (FM) domain structure for perpendicular magnetic anisotropy FM films [27]. As shown in previous studies, $B_c$ could increase as thickness decreases, as observed in van der Waals ferromagnet $Fe_3GeTe_2$, FePd and FePt films [28-30]. While, the opposite behavior that $B_c$ decreases as thickness decreases is also reported, i.e., in Fe and CoFeB thin films [31,32]. Indeed, to fully understand this feature, future experimental verification of the domain structures for various thickness $Fe_{0.29}TaS_2$ flakes would be essential [33,34]. Nevertheless, the strong thickness dependence of $B_c$ in 2D $Fe_{0.29}TaS_2$ could be useful for designing the giant and tunneling magnetoresistance devices based on its heterostructures [16,35-38].

Next, the AHE mechanisms in the quasi-2D van der Waals ferromagnet $Fe_{0.29}TaS_2$ are investigated by studying the scaling behaviors between the anomalous Hall and channel conductivities. Based on the AHE measurements on the 14-nm $Fe_{0.29}TaS_2$ device (Figs. 1(c-f)), the temperature-dependent anomalous Hall resistivities from $T = 2$ to 50 K are calculated, as shown in Fig. 3(a). As the temperature increases, the enhancement of channel resistivity is observed (Fig. 3(b)), which could be associated with stronger phonon and magnon scattering at elevated temperatures. Both the anomalous Hall and channel resistivities decrease as the temperature decreases. Similar features have also been observed on all the other $Fe_{0.29}TaS_2$ devices with various thicknesses. Figs. 3(c) and 3(d) show the temperature-dependent anomalous Hall and channel resistivities measured on the bulk single crystal $Fe_{0.29}TaS_2$ device.

**IV. DISCUSSION**

Decades of theoretical and experimental studies on various FM materials have revealed three AHE mechanisms [18,39], namely skew scattering that is due to the effective spin-orbit coupling of the electrons or the impurities [40,41], side jump mechanism that is due to the electrons that are deflected by impurities [42] and intrinsic contribution that is related to the Berry curvature [43-45]. The skew scattering, intrinsic, and side jump mechanisms give rise to different power law relationship of anomalous Hall conductivity ($\sigma_{AHE}$) and channel conductivity ($\sigma_{xx}$) [18].

$$\sigma_{AHE} \approx \sigma_{xx}^{\gamma} \qquad (1)$$

where $\gamma$ is the power law coefficient. For skew scattering, $\sigma_{AHE}$ is proportional to $\sigma_{xx}$ ($\gamma = 1$), while for intrinsic and side jump mechanisms, $\sigma_{AHE}$ exhibits a constant value as $\sigma_{xx}$ varies ($\gamma = 0$).

To determine the AHE mechanisms in the $Fe_{0.29}TaS_2$, we quantitatively investigate the scaling between $\sigma_{AHE}$ and $\sigma_{xx}$ by analyzing our experimental results using the following formula that takes account of the contributions from both the skew scattering and intrinsic mechanisms:

$$\sigma_{AHE} = \alpha(M)\sigma_{xx} + \beta(M) \qquad (2)$$

where $M$ is the saturation magnetization, and $\alpha(M)$ and $\beta(M)$ represent the skew scattering and the intrinsic contributions to the total anomalous Hall conductivity. It is known in previous studies that the skew scattering contribution term, $\alpha(M)$, is linearly proportional to the saturation magnetization [46]. In Fe$_{0.29}$TaS$_2$, the magnetization decrease is due to low-frequency fluctuations as the temperature varies ($M(T) \approx CT^2$, where C is a constant, see Fig. S5 in Supplemental Material [25]). In this case, the intrinsic contribution term, $\beta(M)$, is expected to also linearly depends on the magnetization [47]. Hence, $\sigma_{AHE}$ can be normalized by the temperature-dependent magnetization ($\sigma_{AHE}^* = \sigma_{AHE} * M(2\text{ K})/M(T)$). It is noted that similar $T_C$ of $\approx 80$ K are obtained for 14-nm, 24-nm and bulk single-crystal Fe$_{0.29}$TaS$_2$ devices (Fig. S6 in Supplemental Material [25]), consistent with other 2D ferromagnets with similar $T_C$ when the thickness is larger than approximately ten layers, such as 2D Fe$_3$GeTe$_2$ and Cr$_2$Ge$_2$Te$_6$ [14,48]. The results of $\sigma_{AHE}^*$ vs $\sigma_{xx}$ measured on various thickness Fe$_{0.29}$TaS$_2$ devices are summarized in Fig. 4 and Fig. S7 in Supplemental Material [25]. Interestingly, $\sigma_{AHE}^*$ vs $\sigma_{xx}$ curves exhibit very different relationships as the Fe$_{0.29}$TaS$_2$ thickness changes. For the 14-nm Fe$_{0.29}$TaS$_2$ device [Fig. 4(a)], $\sigma_{AHE}^*$ exhibit little variation as $\sigma_{xx}$ changes from $3.7 \times 10^3$ to $5.4 \times 10^3$ $\Omega^{-1}$ $cm^{-1}$, which is consistent with intrinsic AHE mechanism arising from the Berry curvature. While for the bulk Fe$_{0.29}$TaS$_2$ device (Fig. 4(d)), $\sigma_{AHE}^*$ increases linearly as $\sigma_{xx}$ increases, which indicates that the skew scattering is the important AHE mechanism. Fig. 4(b) and 4(c) show the $\sigma_{AHE}^*$ vs $\sigma_{xx}$ curves for the 24- and 124-nm Fe$_{0.29}$TaS$_2$ devices. Clearly, the contribution from the skew scattering gradually increases as the Fe$_{0.29}$TaS$_2$ thickness increases. These observations indicate that the thickness variation in van der Waals ferromagnet Fe$_{0.29}$TaS$_2$ could give rise to totally different AHE mechanisms, which makes it an interesting material system for investigating the rich physics of AHE.

To quantitatively separate the skew scattering and intrinsic contributions, the experimental results are numerically fitted based on the following expression:

$$\sigma^*_{AHE} = \alpha_0 \sigma_{xx} + \beta_0 \qquad (3)$$

The red lines in Figs. 4(a)-4(d) represent the best fitting results. Firstly, $\alpha_0$ is obtained to be $(-0.7 \pm 0.6) \times 10^{-3}$ for the 14-nm $Fe_{0.29}TaS_2$ device, which is negligible and further indicates the small contribution from skew scattering. Second, $\alpha_0$ increases as the $Fe_{0.29}TaS_2$ thickness increases, and is obtained to be $(53.2 \pm 0.7) \times 10^{-3}$ for the bulk $Fe_{0.29}TaS_2$ device, indicating the strong enhancement of the skew scattering contribution to the total AHE. The thickness dependence of skew scattering parameter $\alpha_0$ is summarized in Fig. S8 in Supplemental Material [25]. The intrinsic contribution parameters $\beta_0$ for $Fe_{0.29}TaS_2$ bulk single crystal and 14nm flake are obtained to be $(119.7 \pm 13.8)$ and $(43.2 \pm 2.5)$ $\Omega^{-1}\ cm^{-1}$, respectively. The values are within the range of previous theoretical and experimental results of ferromagnetic metals and semiconductors [18,39,49]. Compared to magnetic Weyl semimetal candidates $Fe_{3-x}GeTe_2$ and $Co_3Sn_2S_2$ [50,51], these intrinsic values are significantly smaller, indicating the smaller Berry phase-induced effective spin-orbit coupling in $Fe_{0.29}TaS_2$. $\beta_0$ exhibits some variation as the $Fe_{0.29}TaS_2$ thickness changes, which might be due to different Berry curvature associated with different Fermi level positions evidenced by thickness-dependent carrier density (Fig. S9 in Supplemental Material [25]), as shown in previous reports on (Ge, Mn)Te and $Mn_{1.5}Ga$ [52,53]. Hence, the intrinsic anomalous Hall mechanism provides a unique experimental method to probe the Berry curvature-related physics for the emergent 2D ferromagnetic materials [8,9].

To investigate the crossover from the intrinsic to skew scattering mechanisms as the $Fe_{0.29}TaS_2$ thickness increases, we plot the ratio of the intrinsic-contributed anomalous Hall conductivity over the total normalized anomalous Hall conductivity ($\beta_0/\sigma^*_{AHE}$) as a function of the temperature and $Fe_{0.29}TaS_2$ thickness (Fig. 5(a)). It is noticed that skew scattering plays a major role in thicker $Fe_{0.29}TaS_2$. Consistent with previous studies of AHE, high conductivity favors skew scattering contribution and intrinsic contribution is more relevant in the low conductivity region [18,39]. As the thickness of $Fe_{0.29}TaS_2$ decreases, the enhanced surface and defect scatterings decrease the scattering time, leading to less contribution from skew scattering. Similarly, as temperature increases, the increased magnon and phonon scatterings also decrease the scattering time and the channel conductivity, which also lead to less skew scattering contribution. Clearly, there is a strong correlation between the anomalous Hall mechanisms and the channel conductivities as a function of the temperature and $Fe_{0.29}TaS_2$ thickness [Fig. 5(a) and 5(b)].

Figure 6 summarizes all the results of $\sigma^*_{AHE}$ vs $\sigma_{xx}$ measured on all the $Fe_{0.29}TaS_2$ devices with various thicknesses. In the low conductivity region, $\sigma^*_{AHE}$ exhibits little variation as $\sigma_{xx}$ changes, while in the high conductivity region, $\sigma^*_{AHE}$ exhibits a linearly relationship as $\sigma_{xx}$ increases. The shadowed area indicates the crossover of the major AHE mechanism between the extrinsic skew scattering and the intrinsic contribution with $\sigma_{xx} \approx 8 \times 10^3 \, \Omega^{-1} \, cm^{-1}$. As demonstrated in previous theoretical study, it is shown that the major anomalous Hall mechanism is related to the spin orbit coupling energy ($E_{soc}$) and the electron scattering time ($\tau$) in the ferromagnetic materials [39]. In the case for $E_{soc} \gg \frac{h}{\tau}$ ($h$ is the Planck constant), skew scattering plays the dominate role. Hence, in the region of channel conductivity, $\sigma_{xx} \gg \frac{ne^2h}{mE_{soc}}$ (*m* is the effective electron mass, *n* is the carrier density, and *e* is the electron charge), skew scattering is the dominant AHE mechanism. When the conductivity decreases and becomes smaller than $\frac{ne^2h}{mE_{soc}}$, the AHE mechanism changes from the extrinsic skew scattering to the intrinsic mechanism. The critical channel conductivity in $Fe_{0.29}TaS_2$ devices for the crossover of AHE mechanisms is $\approx 8 \times 10^3 \, \Omega^{-1} \, cm^{-1}$, which is smaller compared to the values for conventional ferromagnetic metals of $\approx 10^5$ to $10^6 \, \Omega^{-1} \, cm^{-1}$ [39]. This might be associated with the lower carrier density in $Fe_{0.29}TaS_2$ devices ($\approx 10^{20}$ cm$^{-3}$) compared to conventional ferromagnetic metals ($\approx 10^{23}$ cm$^{-3}$) and different spin orbit coupling energy.

## V. CONCLUSION

In summary, the AHE mechanisms in quasi-2D van der Waals ferromagnet $Fe_{0.29}TaS_2$ are investigated via the scaling relationship between the anomalous Hall and channel conductivities. The dominant role for AHE in bulk $Fe_{0.29}TaS_2$ is determined to be skew scattering, while as the $Fe_{0.29}TaS_2$ thickness decreases, the contribution from the intrinsic mechanism increases and becomes dominant for the 14-nm $Fe_{0.29}TaS_2$ device. This crossover is found to be highly associated with the thickness-dependent channel conductivity in $Fe_{0.29}TaS_2$. Additionally, $B_c$ decreases as the $Fe_{0.29}TaS_2$ thickness decreases, which could be associated with variation of the FM domain structure. Our results pave the way for further understanding of fundamental magnetic properties in van der Waals ferromagnets and could be useful for future 2D materials-based spintronics devices.

**ACKNOWLEDGEMENTS**

We acknowledge the financial support from National Basic Research Programs of China (973 program Grant Nos. 2015CB921104 and 2018YFA0305601), National Natural Science Foundation of China (NSFC Grant No. 11574006, No. 11774007 and No. U1832214), Beijing Natural Science Foundation (No. 1192009), and the Key Research Program of the Chinese Academy of Sciences (Grant No. XDB28000000).

[46] P. Nozières and C. J. J. P. F. Lewiner, A simple theory of the anomalous hall effect in semiconductors. *J. Phys. France* **34**, 901 (1973).

[47] C. Zeng, Y. Yao, Q. Niu, and H. H. Weitering, Linear Magnetization Dependence of the Intrinsic Anomalous Hall Effect. *Phys. Rev. Lett.* **96**, 037204 (2006).

[48] M. Kim, P. Kumaravadivel, J. Birkbeck, W. Kuang, S. G. Xu, D. G. Hopkinson, J. Knolle, P. A. McClarty, A. I. Berdyugin, M. B. Shalom, R. V. Gorbachev, S. J. Haigh, S. Liu, J. H. Edgar, K. S. Novoselov, I. V. Grigorieva, and A. K. Geim, Hall micromagnetometry of individual two-dimensional ferromagnets. *arXiv:1902.06988* (2019).

[49] D. Hou, G. Su, Y. Tian, X. Jin, S. A. Yang, and Q. Niu, Multivariable Scaling for the Anomalous Hall Effect. *Phys. Rev. Lett.* **114**, 217203 (2015).

[50] E. Liu, Y. Sun, N. Kumar, L. Muechler, A. Sun, L. Jiao, S.-Y. Yang, D. Liu, A. Liang, Q. Xu, J. Kroder, V. Süß, H. Borrmann, C. Shekhar, Z. Wang, C. Xi, W. Wang, W. Schnelle, S. Wirth, Y. Chen, S. T. B. Goennenwein, and C. Felser, Giant anomalous Hall effect in a ferromagnetic kagome-lattice semimetal. *Nat. Phys.* **14**, 1125 (2018).

[51] K. Kim, J. Seo, E. Lee, K. T. Ko, B. S. Kim, B. G. Jang, J. M. Ok, J. Lee, Y. J. Jo, W. Kang, J. H. Shim, C. Kim, H. W. Yeom, B. Il Min, B.-J. Yang, and J. S. Kim, Large anomalous Hall current induced by topological nodal lines in a ferromagnetic van der Waals semimetal. *Nat. Mater.* **17**, 794 (2018).

[52] R. Yoshimi, K. Yasuda, A. Tsukazaki, K. S. Takahashi, M. Kawasaki, and Y. Tokura, Current-driven magnetization switching in ferromagnetic bulk Rashba semiconductor (Ge,Mn)Te. *Sci. Adv.* **4**, eaat9989 (2018).

[53] L. J. Zhu, D. Pan, and J. H. Zhao, Anomalous Hall effect in epitaxial $L1_0$ Mn$_{1.5}$Ga films with variable chemical ordering. *Phys. Rev. B* **89**, 220406 (2014).

**Figure 1**

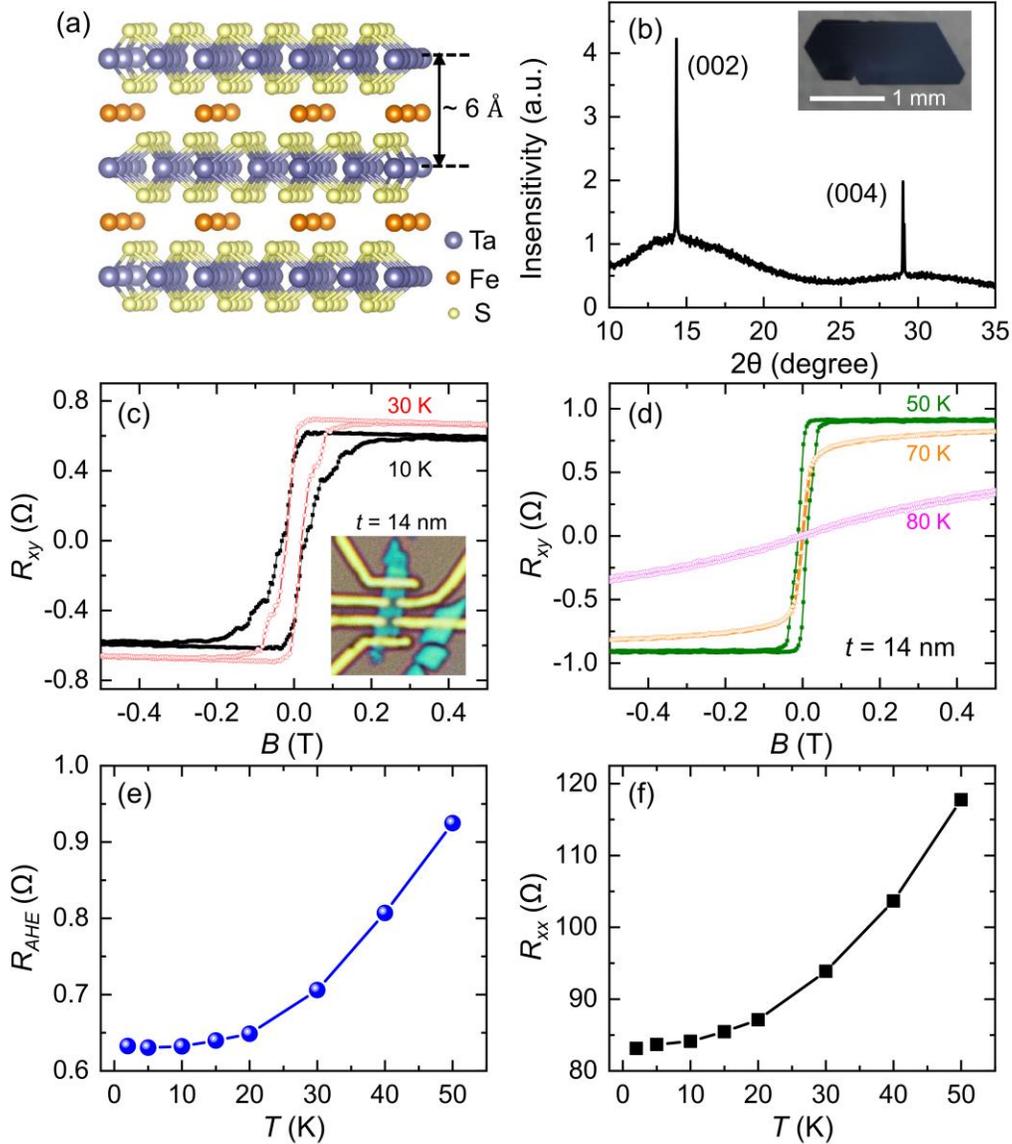

Fig. 1. Crystal structure and Hall measurements of quasi-2D van der Waals ferromagnet $Fe_{0.29}TaS_2$. (a) Crystal structure of layered Fe-intercalated van der Waals ferromagnet $Fe_{0.29}TaS_2$. (b) X-ray-diffraction characterization of the synthesized $Fe_{0.29}TaS_2$ single crystals. Inset: the optical image of a typical synthesized crystal. (c-d) The Hall measurements on a typical 14 nm $Fe_{0.29}TaS_2$ device at 10, 30, 50, 70, and 80 K, respectively. Inset: the optical image of this $Fe_{0.29}TaS_2$ device. (e-f) The temperature dependence of the anomalous Hall resistance ($R_{AHE}$) and the channel resistance ($R_{xx}$) measured on this device.

**Figure 2**

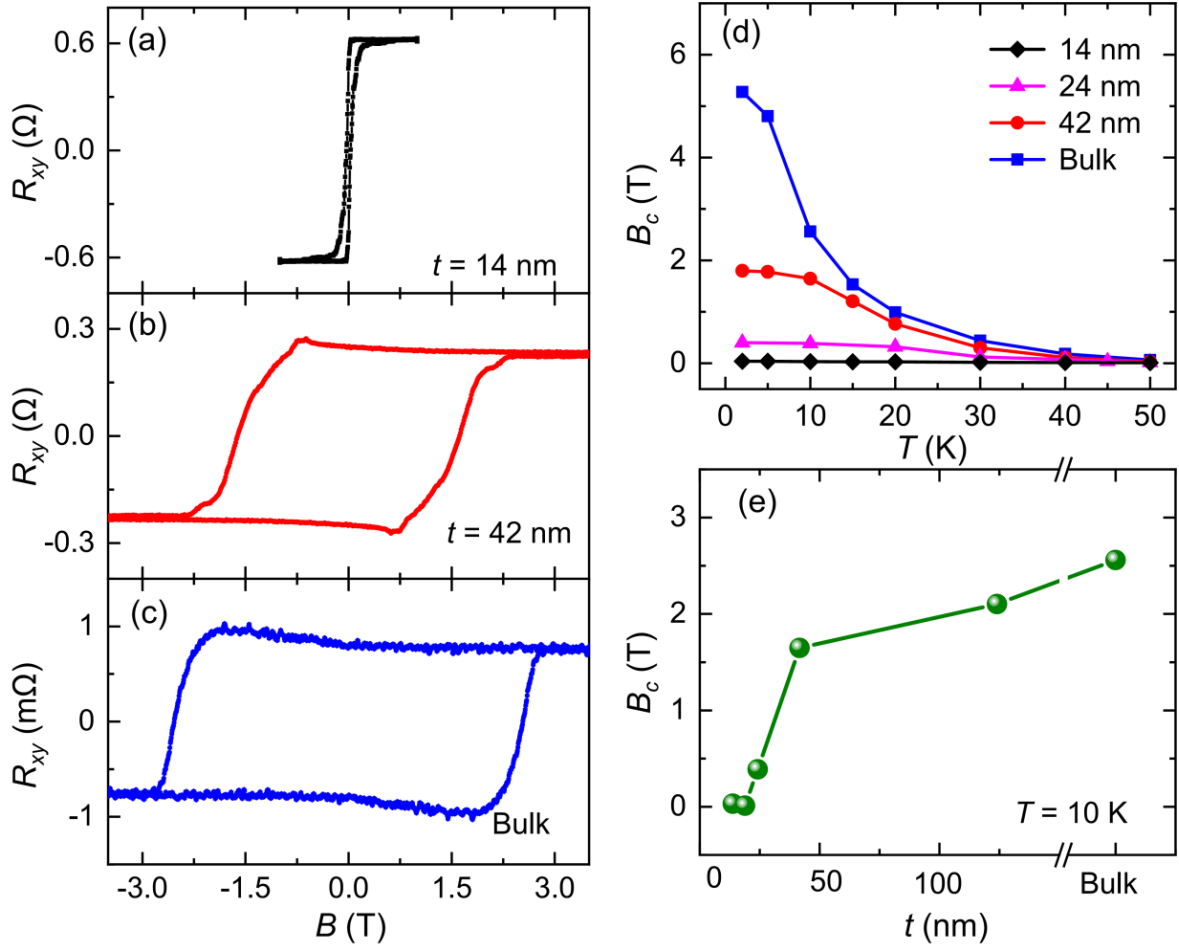

Fig. 2. The coercive magnetic field ($B_c$) probed via AHE measurements. (a-c) The AHE results measured on 14-nm, 42-nm, and bulk $Fe_{0.29}TaS_2$ devices, respectively, at $T = 10$ K. (d) $B_c$ as a function of temperature for $Fe_{0.29}TaS_2$ devices with thickness of 14-nm, 24-nm, 42-nm, and the bulk single-crystal device. (e) the thickness dependence of $B_c$ at $T = 10$ K.

**Figure 3**

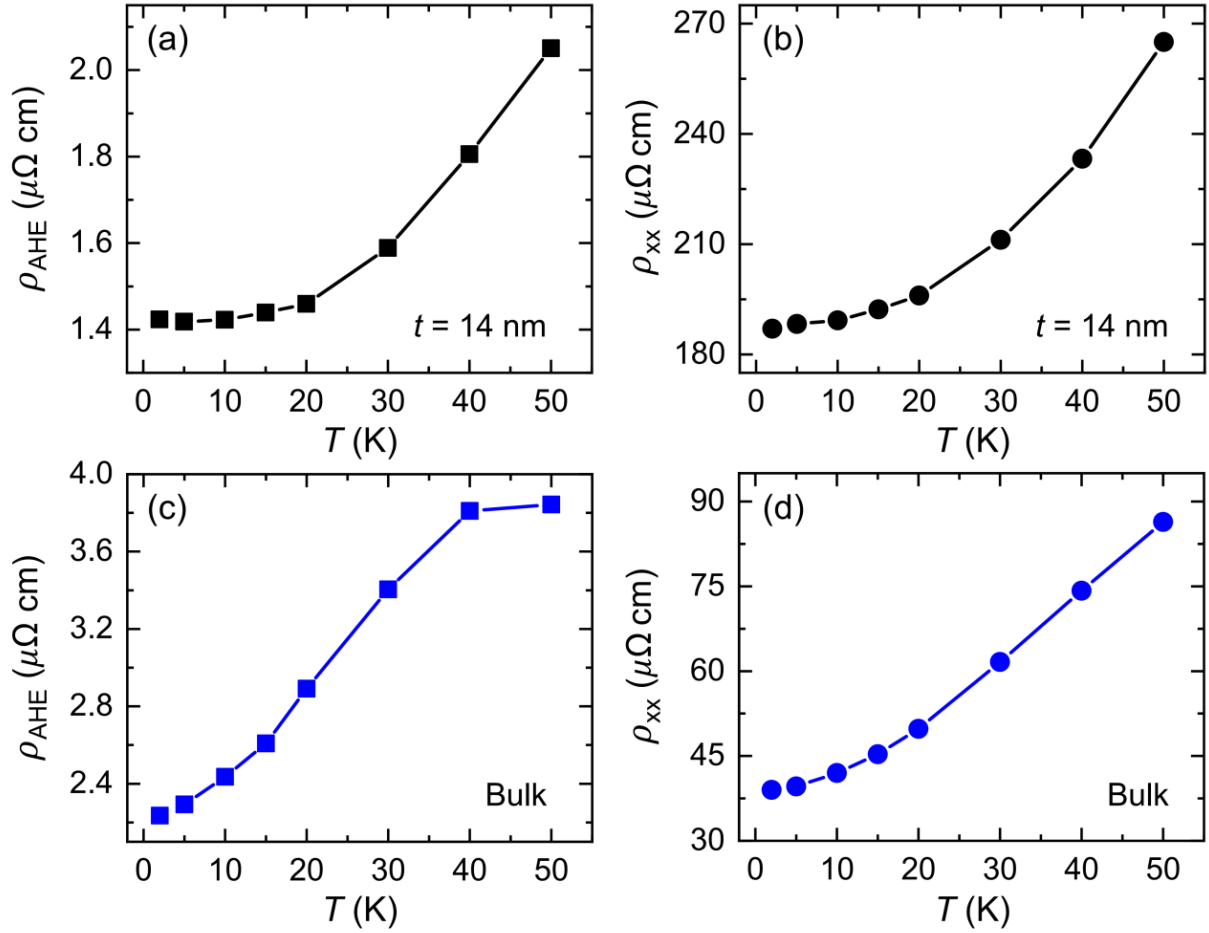

Fig. 3. The anomalous Hall and channel resistivities as a function of the temperature. (a-b) The temperature dependence of the anomalous Hall and channel resistivities for the 14-nm $Fe_{0.29}TaS_2$ device. (c-d) The temperature dependence of the anomalous Hall and channel resistivities for the bulk $Fe_{0.29}TaS_2$ device.

**Figure 4**

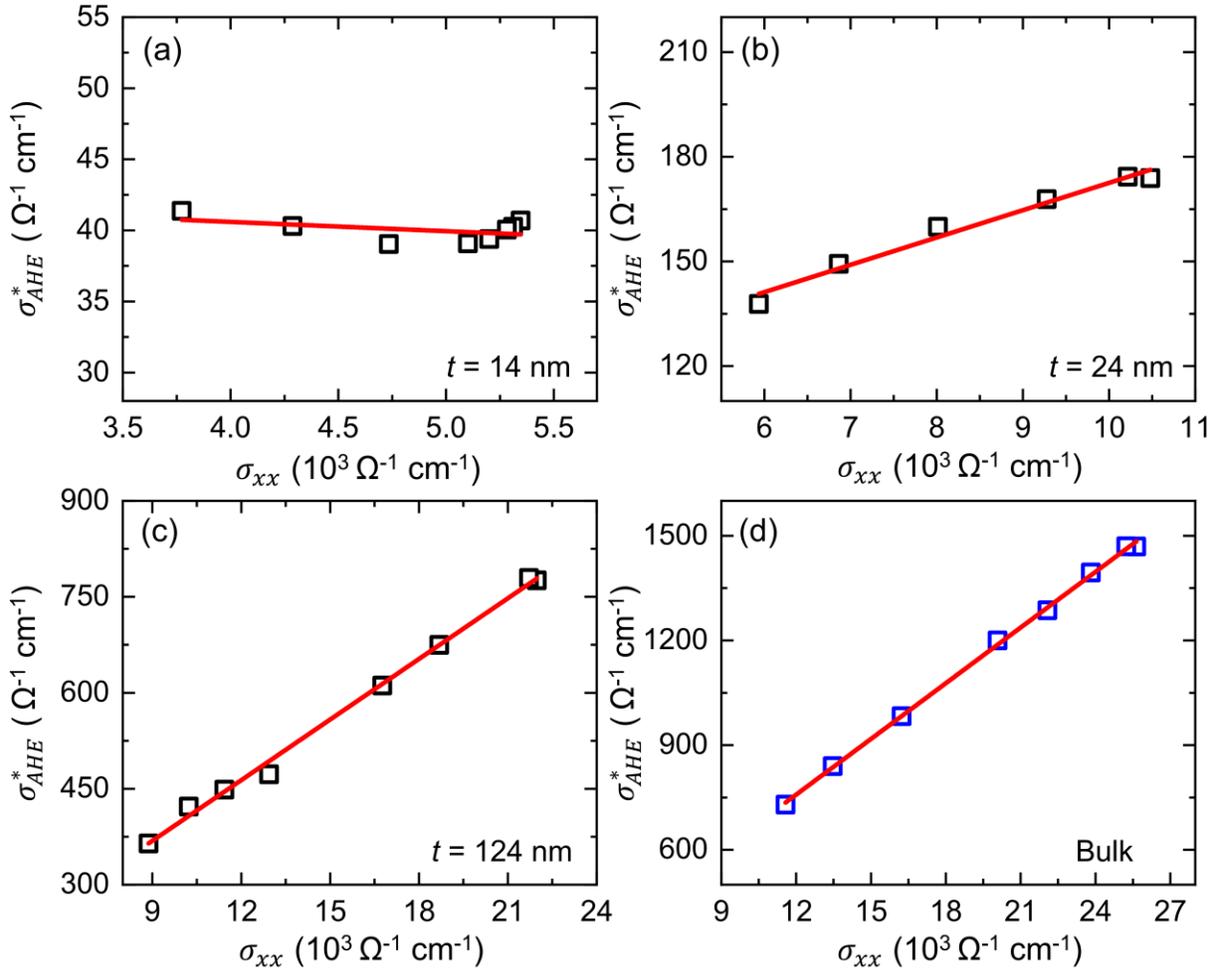

Fig. 4. Investigation of the AHE mechanisms in $Fe_{0.29}TaS_2$ via scaling relationship between anomalous Hall and channel conductivities for 14-nm $Fe_{0.29}TaS_2$ (a), 24-nm $Fe_{0.29}TaS_2$ (b), 124-nm $Fe_{0.29}TaS_2$ (c), and bulk $Fe_{0.29}TaS_2$ (d), respectively.

**Figure 5**

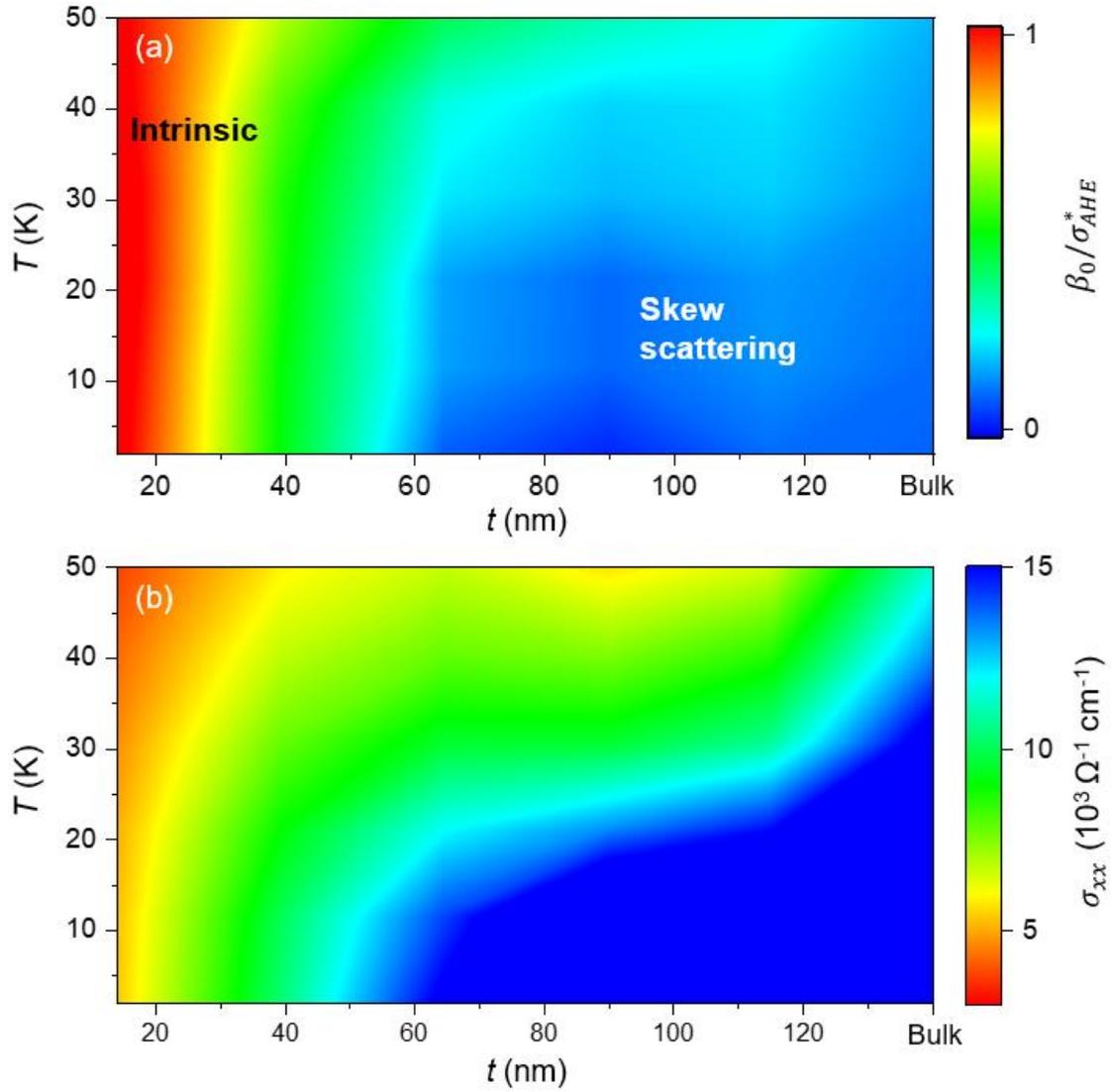

Fig. 5. The strong correlation of the anomalous Hall mechanism and the channel conductivity. (a) The temperature and thickness dependences of the intrinsic contribution and skew scattering to the total anomalous Hall conductivity. (b) The temperature and thickness dependences of the channel conductivity of $Fe_{0.29}TaS_2$.

**Figure 6**

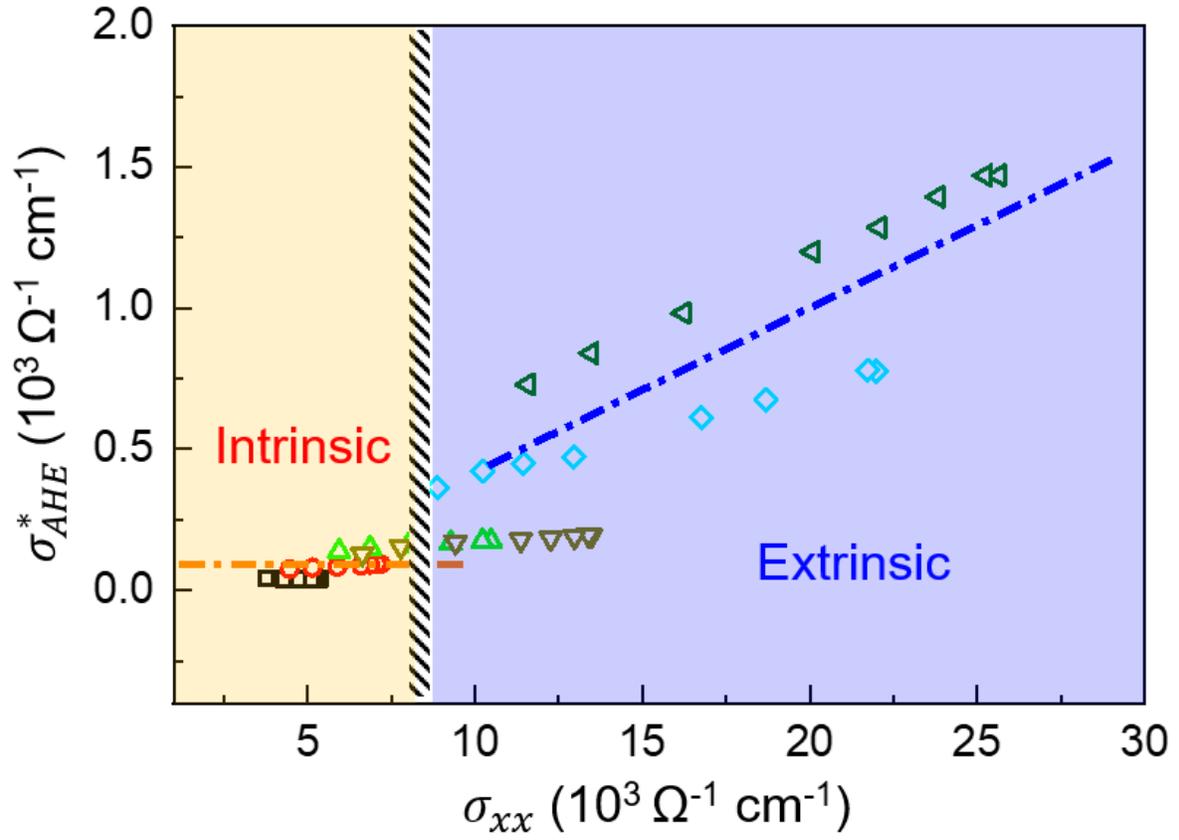

Fig. 6. The crossover from extrinsic skew scattering to intrinsic contribution as the channel conductivity of $Fe_{0.29}TaS_2$ decreases.

**Supplementary Materials for:**

# Anomalous Hall Effect Mechanisms in Quasi-2D van der Waals Ferromagnet Fe$_{0.29}$TaS$_2$

Ranran Cai[1,2], Wenyu Xing[1,2], Huibin Zhou[1,2], Boning Li[1,2], Yangyang Chen[1,2], Yunyan Yao[1,2], Yang Ma[1,2], X. C. Xie[1,2,3,4], Shuang Jia[1,2,3,4], and Wei Han[1,2*]

[1] International Center for Quantum Materials, School of Physics, Peking University, Beijing 100871, P. R. China

[2] Collaborative Innovation Center of Quantum Matter, Beijing 100871, P. R. China

[3] CAS Center for Excellence in Topological Quantum Computation, University of Chinese Academy of Sciences, Beijing 100190, P. R. China

[4] Beijing Academy of Quantum Information Sciences, Beijing 100193, P. R. China

* Correspondence to: weihan@pku.edu.cn

**Figure S1**

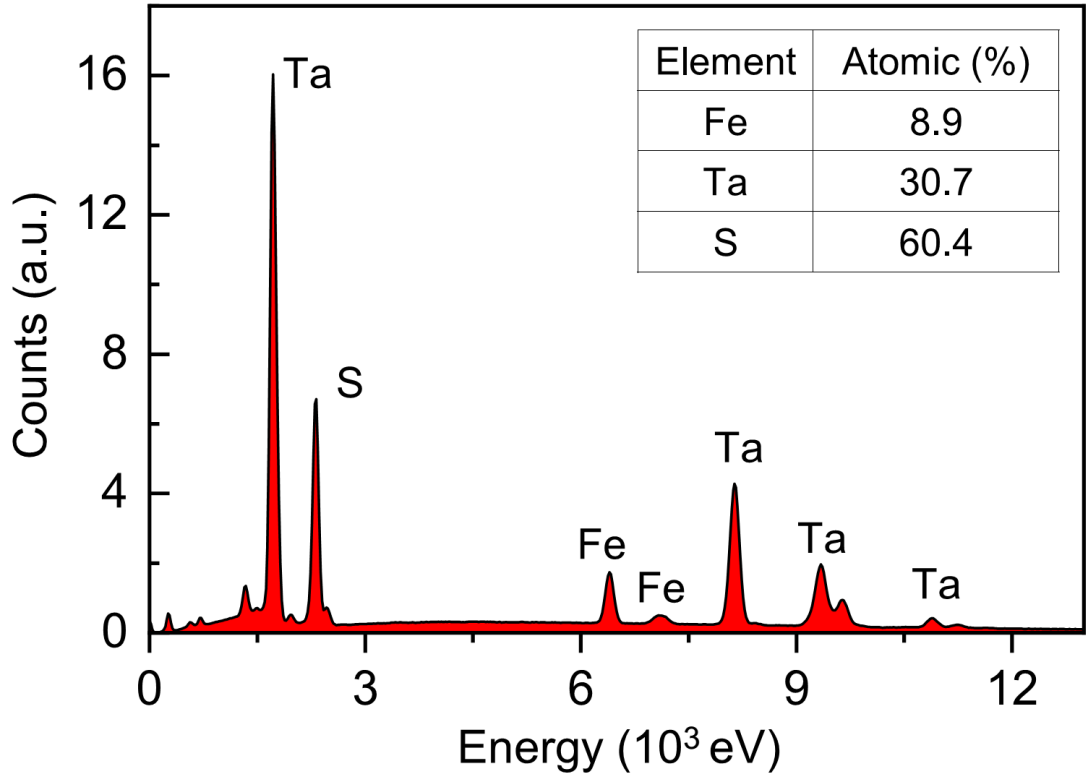

Fig. S1. The energy-dispersive spectroscopy of synthesized Fe$_x$TaS$_2$ bulk single crystals.

**Figure S2**

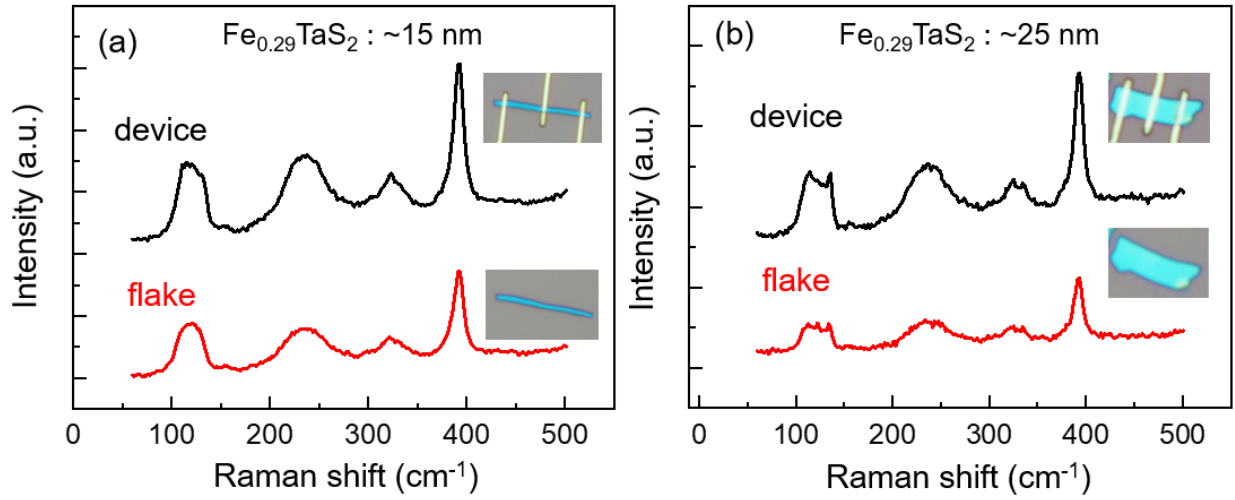

Fig. S2. The Raman spectra of pristine $Fe_{0.29}TaS_2$ flakes and the devices with thickness of ~15 nm (a) and ~25 nm (b).

**Figure S3**

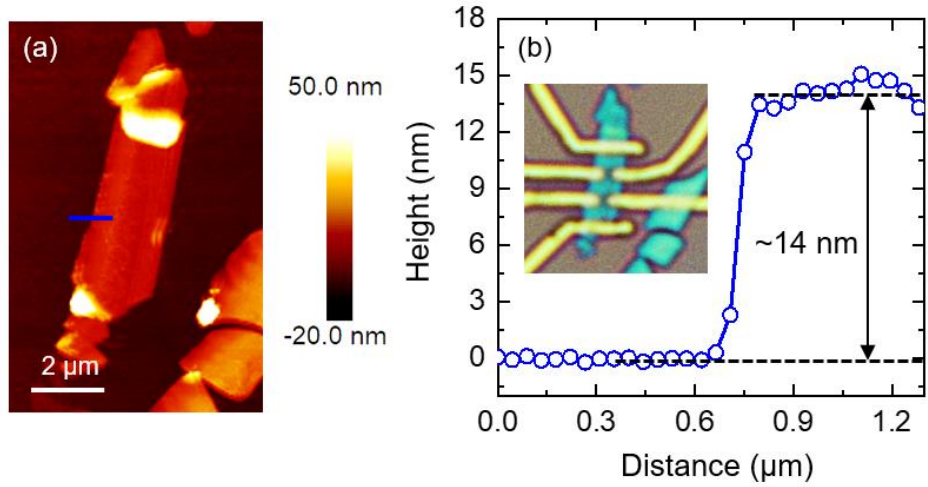

Fig. S3. Measurement of the $Fe_{0.29}TaS_2$ thin flake thickness via atomic force microscopy (AFM). (a) the AFM image of the 14 nm $Fe_{0.29}TaS_2$ thin flake. (b) The height profile across the $Fe_{0.29}TaS_2$ flake along the blue dashed line in (a). Inset: optical image of 14 nm $Fe_{0.29}TaS_2$ device.

**Figure S4**

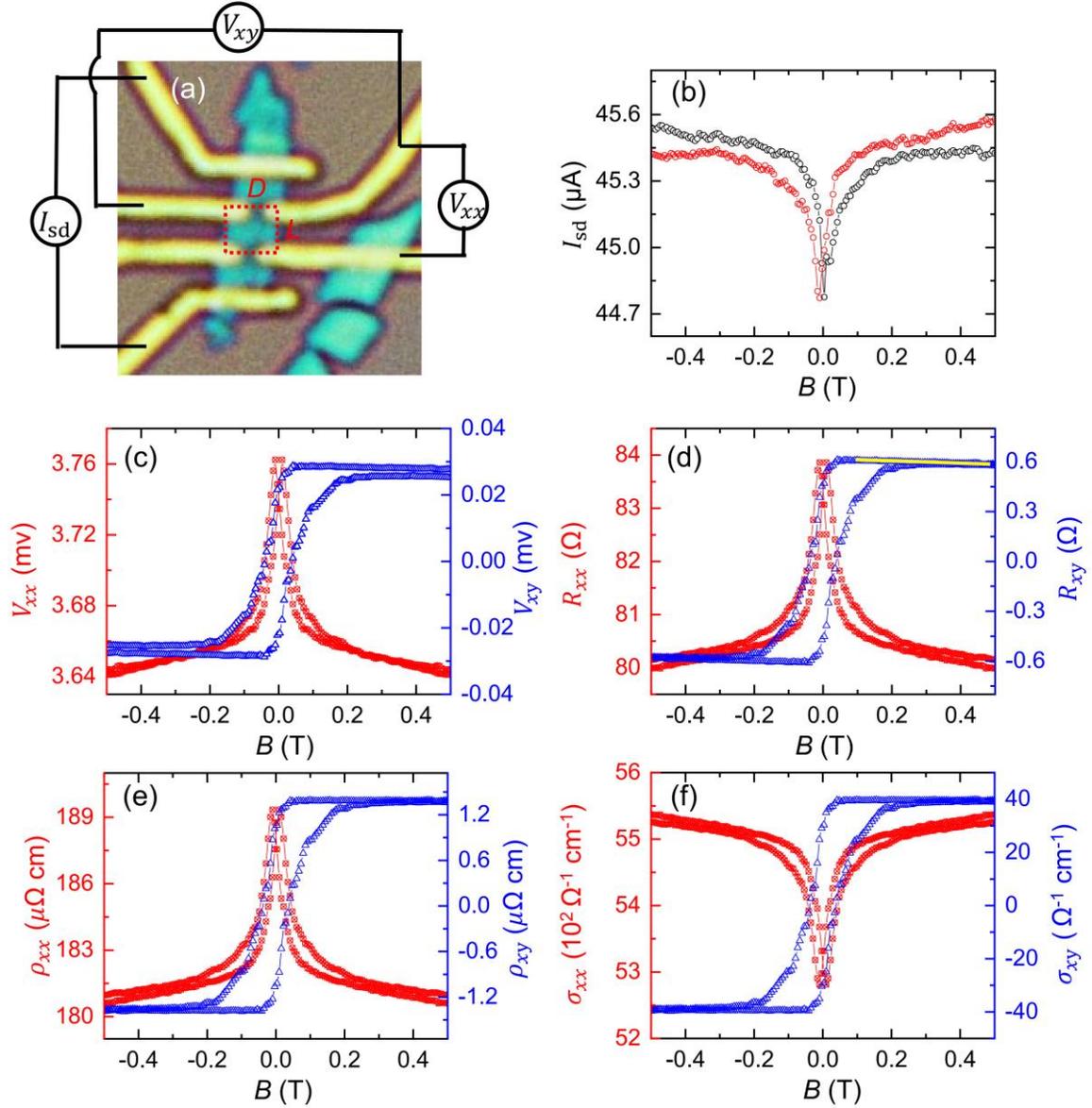

Fig. S4. The data analysis procedures of the typical device made on 14 nm $Fe_{0.29}TaS_2$. (a) Hall bar measurement geometry on $Fe_{0.29}TaS_2$ with channel length (*L*) and width (*D*) (red dashed square). (b-c) The current from source to drain ($I_{sd}$) and the channel and Hall voltages ($V_{xx}$ and $V_{xy}$) as a function of perpendicular magnetic field. (d) Magnetic field dependent channel resistance ($R_{xx}$ =

$\frac{V_{xx}}{I_{sd}}$) and Hall resistance ($R_{xy} = \frac{V_{xy}}{I_{sd}}$) calculate from (b-c). (e) Magnetic field dependent channel resistivity ($\rho_{xx} = R_{xx}\frac{D \cdot t}{L}$) and Hall resistivity ($\rho_{xy} = R_{xy}\frac{D \cdot t}{L}$) calculated from (d). (f) Magnetic field dependent channel conductivity ($\sigma_{xx} = \frac{\rho_{xx}}{\rho_{xx}^2 + \rho_{xy}^2} \sim \frac{1}{\rho_{xx}}$, for $\rho_{xx} \gg \rho_{xy}$) and Hall resistivity ($\sigma_{xy} = \frac{\rho_{xy}}{\rho_{xx}^2 + \rho_{xy}^2} \sim \frac{\rho_{xy}}{\rho_{xx}^2}$, for $\rho_{xx} \gg \rho_{xy}$) calculated from (e).

**Figure S5**

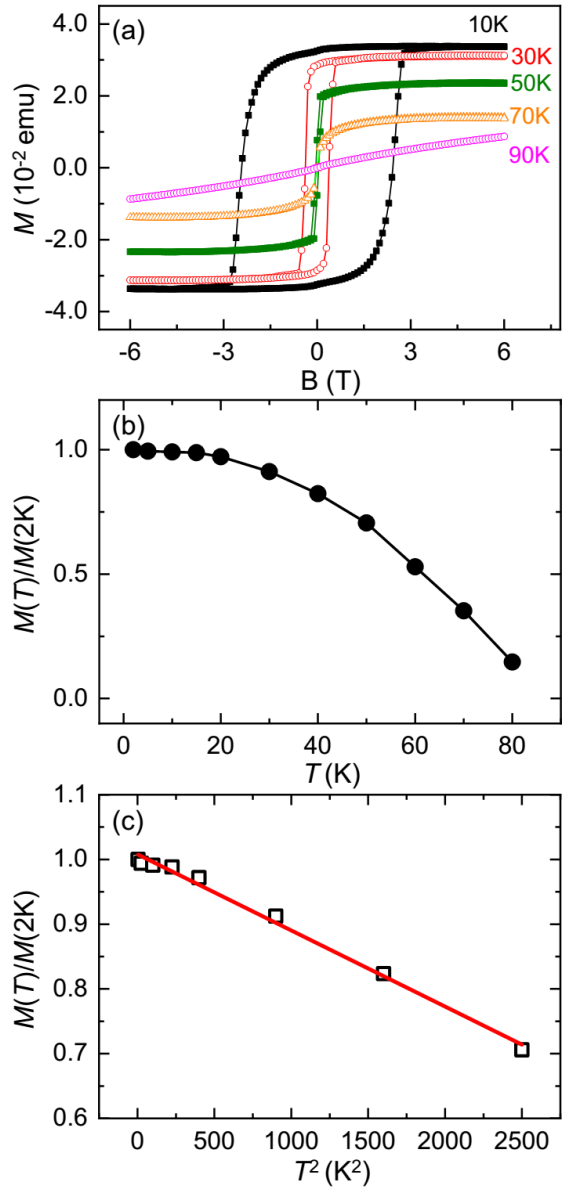

Fig. S5. The temperature dependence of magnetization (*M*) in $Fe_{0.29}TaS_2$ bulk single crystal. (a) Magnetization curves at *T* = 10K, 30K, 50K, 70K, and 90K, respectively. (b) Temperature dependence of *M*. (c) *M* versus $T^2$ from 2K to 50K, the solid red line is the best linear fitting of experimental data (open black dots). The $T^2$ dependence of magnetization is related to the long-wavelength, low-frequency fluctuations.

**Figure S6**

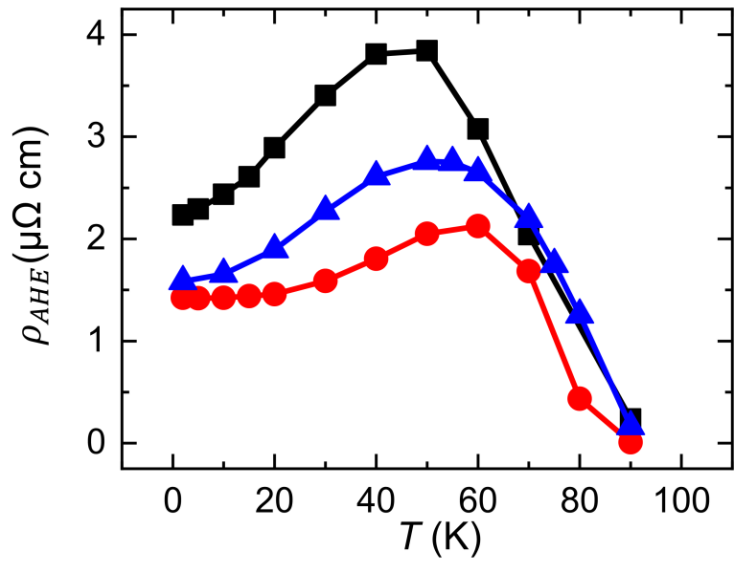

Fig. S6. Temperature dependent AHE resistivity for 14nm (red), 24nm (blue) and bulk (black) devices. Similar Curie temperatures of ~ 80 K are obtained as the thickness of $Fe_{0.29}TaS_2$ decreases from bulk to 14 nm (~ 23 layers).

**Figure S7**

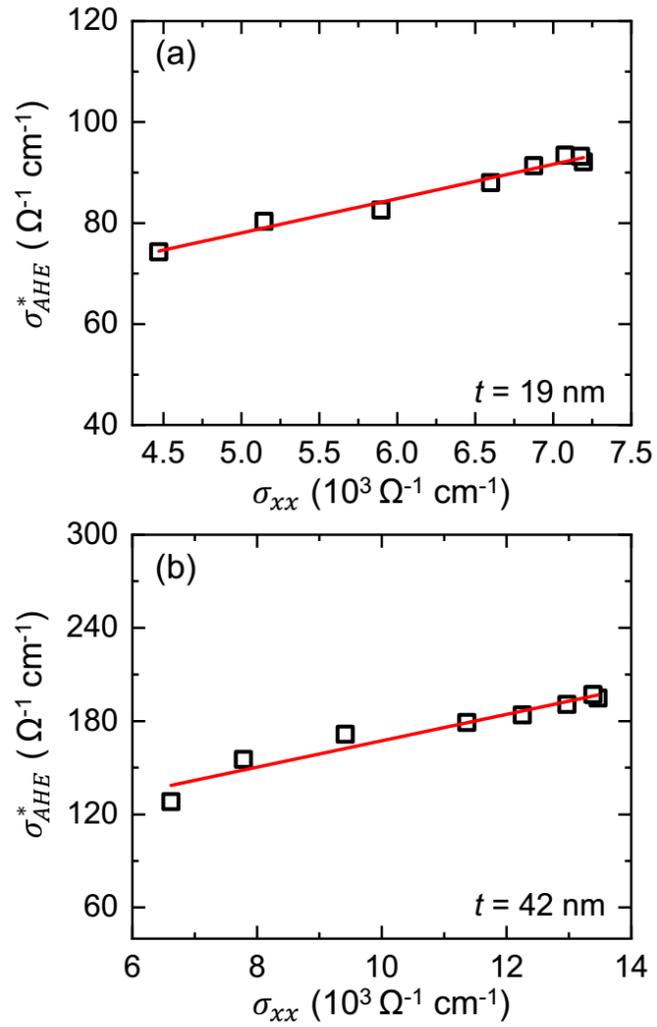

Fig. S7. The scaling relationship between normalized anomalous Hall and channel conductivities for 19 nm $Fe_{0.29}TaS_2$ (a) and 42 nm $Fe_{0.29}TaS_2$ (b) devices.

**Figure S8**

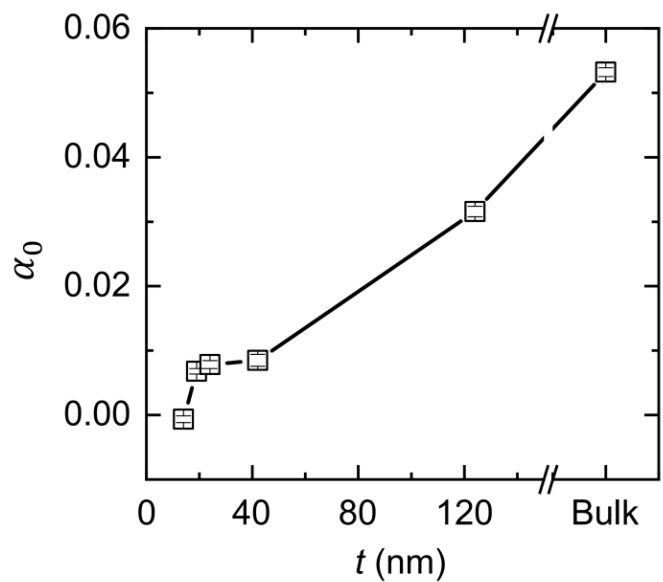

Fig. S8. The $Fe_{0.29}TaS_2$ thickness dependence of skew scattering parameter $\alpha_0$.

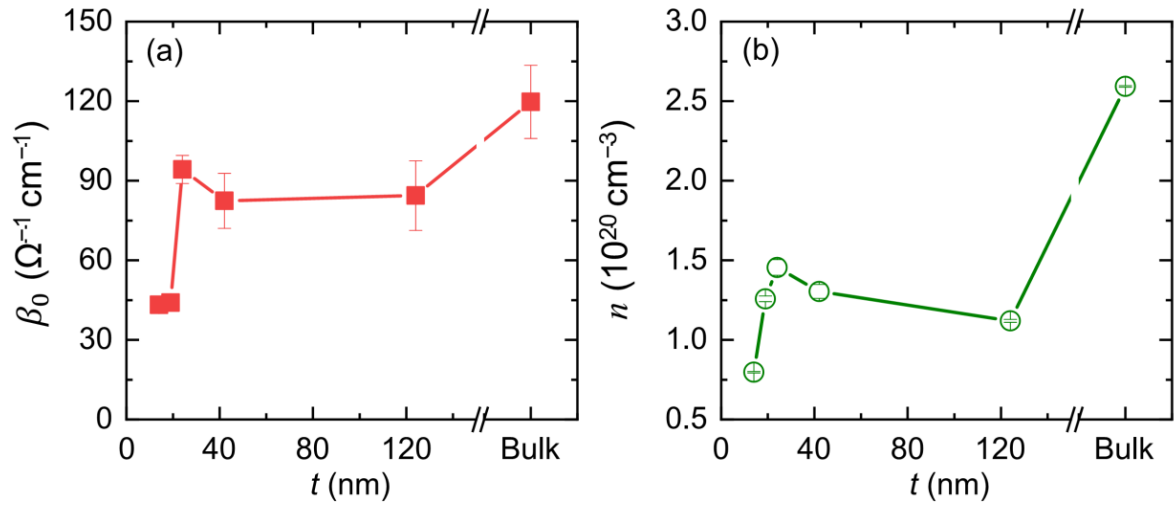

Fig. S9. $Fe_{0.29}TaS_2$ thickness dependence of the intrinsic anomalous conductivity $\beta_0$ (a) and the carrier density (b).

**Figure S10**

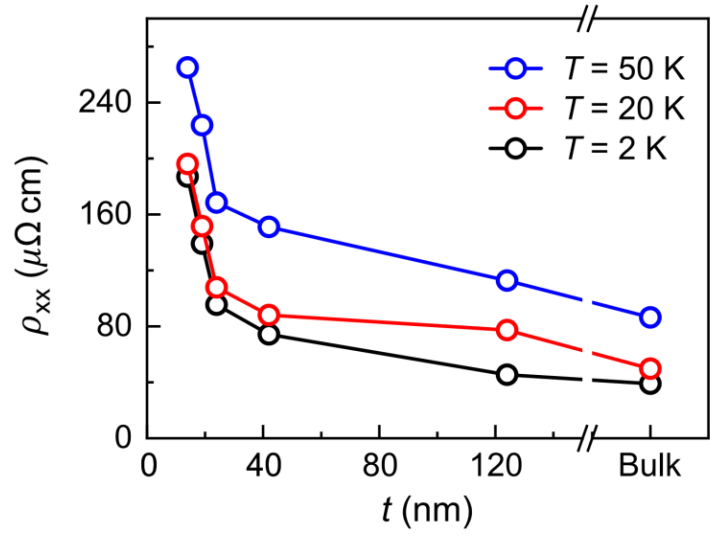

Fig. S10. Thickness dependence of $Fe_{0.29}TaS_2$ channel resistivity at $T$ = 2, 20 and 50 K, respectively.